\documentclass[aps,pre,reprint,superscriptaddress,nofootinbib]{revtex4-2}
\usepackage{graphicx,epstopdf,amssymb,amsmath,verbatim,hyperref,mathdots,xcolor}
\usepackage{tikz}
\pdfoutput=1

\hypersetup{colorlinks=true,citecolor=blue,
linkcolor=blue,urlcolor=blue,pdfstartview=FitH,
bookmarksopen=true}

\begin{document}

\title{Directed Polymer Transfer Matrices as a Unified Generator of Distinct One-Point Fluctuation Laws
}

\author{Sen Mu}
\affiliation{Max Planck Institute for the Physics of Complex Systems, 01187 Dresden, Germany}

\author{Abbas Ali Saberi}\email{asaberi@constructor.university}
\affiliation{School of Science, Constructor University, Campus Ring 1, 28759 Bremen,
Germany}\affiliation{Max Planck Institute for the Physics of Complex Systems, 01187 Dresden, Germany}

\author{Roderich Moessner}
		\affiliation{Max Planck Institute for the Physics of Complex Systems, 01187 Dresden, Germany}

\author{Mehran Kardar}
\affiliation{Department of Physics, Massachusetts Institute of Technology, Cambridge, USA}
\date{\today}

\begin{abstract}

We numerically revisit the transfer-matrix formulation of directed polymers in random media and show that a common finite-dimensional framework organizes the canonical one-point fluctuation laws in $(1+1)$ dimensions. For a fixed realization of the bulk disorder, full-space partition functions are obtained from the same time-ordered product $W(t)$ through endpoint contractions or a Brownian-weighted initial vector, while the half-space construction modifies only the transfer rule at the absorbing boundary. These choices yield distributions consistent with the standard KPZ subclasses: Tracy--Widom GUE for point-to-point geometry, Tracy--Widom GOE for point-to-line geometry, Tracy--Widom GSE for half-space point-to-point geometry, and Baik--Rains for the stationary line-to-point construction. In all four cases, the free-energy fluctuations grow as $t^{1/3}$, and the low-order cumulants approach the corresponding universal benchmarks. The matrix-product formulation also provides access to intrinsic spectral observables. For the leading eigenvalue $\lambda_1(t)$, the fluctuations of $\ln\lambda_1(t)$ exhibit an intermediate $t^{1/3}$ regime, while the standardized distribution remains distinct from the canonical benchmark laws over the studied time range.

\end{abstract}

\maketitle

\section{Introduction}
\label{sec:introduction}

The Kardar-Parisi-Zhang (KPZ) universality class governs a broad range of non-equilibrium fluctuation phenomena in one spatial dimension, including stochastic interface growth, interacting particle systems, and directed polymers in random media (DPRM)~\cite{kardar1986dynamic,halpinhealy1995kinetic,kriecherbauer2010pedestrian,corwin2012kpz,takeuchi2018appetizer}.
Beyond model-specific details, KPZ systems share a characteristic scaling structure: For growing interfaces, typical height fluctuations initially grow in time as $t^{\beta}$ with $\beta=1/3$,
spatial correlations spread as $t^{1/z}$ with dynamic exponent $z=3/2$, ultimately saturating at finite size with roughness exponent $\alpha=\beta z$~\cite{kardar1986dynamic,halpinhealy1995kinetic,kriecherbauer2010pedestrian,takeuchi2018appetizer}.
An equivalent formulation of the same universality class is provided by directed polymers in random media (DPRM), in which one studies a directed path propagating through a quenched random energy landscape; the logarithm of its partition function defines a free energy whose fluctuations obey the same $t^{1/3}$  scaling, while the transverse wandering of the polymer endpoint scales as $t^{2/3}$.

A central insight of the modern theory is that KPZ universality in $1+1$ dimensions contains distinct one-point fluctuation subclasses, selected by geometry and boundary conditions (or, for full-line growth, by initial data) \cite{prahofer2000universal,baik2000limiting,corwin2012kpz,takeuchi2018appetizer}. For curved (droplet) geometry, the properly centered and scaled height/free-energy fluctuations converge to the Tracy--Widom law associated with the Gaussian unitary ensemble (GUE) \cite{johansson2000shape,tracywidom1994airy,sasamoto2010exact,amir2011continuum}. For flat geometry, the limiting form is the Tracy--Widom law associated with the Gaussian orthogonal ensemble (GOE) \cite{prahofer2000universal,tracywidom1996orthogonal,calabreseLeDoussal2012flat}. For stationary geometry, obtained by imposing a two-sided Brownian initial profile, the limiting one-point law is the Baik--Rains distribution \cite{baik2000limiting,prahofer2000universal}. In half-space settings, the presence of a boundary introduces additional subclasses; in particular, an absorbing or sufficiently repulsive wall yields the Tracy--Widom law associated with the Gaussian symplectic ensemble (GSE) \cite{sasamoto2004halfspace,imamura2004external,tracywidom1996orthogonal,gueudreLeDoussal2012hardwall}. (Here GUE, GOE, and GSE refer to the random-matrix ensembles whose soft-edge laws appear in the KPZ problem; the transfer-matrix product studied below is not itself sampled from one of these invariant ensembles.) Different KPZ subclasses can compete and lead to crossovers when distinct growth sectors are coupled, as shown for growth on crossing substrates with various subclass pairings \cite{SaberiDashtiNaserabadiKrug2019}. These results provide one of the clearest examples of universal fluctuation laws controlled by geometry rather than microscopic dynamics.

Traditionally, these subclasses are realized through distinct dynamical settings: different initial data, boundary geometries, or half-space constructions are imposed at the level of the growth process itself. In this sense, the canonical one-point laws appear as outcomes of separate evolutions, each tailored to a particular geometry.

The directed-polymer formulation is historically one of the foundational routes into KPZ physics. In the lattice DPRM, the partition function is obtained by summing Boltzmann weights over all directed paths through a quenched random energy landscape, and its logarithm encodes the polymer free energy \cite{kardar1987replica,halpinhealy1995kinetic,kriecherbauer2010pedestrian,corwin2012kpz}. The path sum can be evaluated recursively with a transfer-matrix algorithm, in which the partition-sum vector at time $t$ is obtained from that at time $t-1$ by multiplication with a random near-diagonal matrix. Transfer-matrix products were used in early analytical and numerical studies of directed polymers, including finite-temperature thermodynamics and free-energy probability distributions \cite{derridaGolinelli1990,halpinHealy1991,kardar1987replica,kim1991zerotemperature,halpinhealy1995kinetic}. This formulation makes the growth recursion explicit, gives direct access to finite-time fluctuation statistics, and naturally suggests studying both matrix elements and spectral quantities.

The spectral perspective also has a substantial history. General products of random matrices are governed by Lyapunov growth rates \cite{furstenbergKesten1960,oseledets1968,hennion1997}, and Eckmann and Wayne established an explicit connection between the largest Lyapunov exponent of random matrix products and directed polymers in random environments \cite{eckmannWayne1989}. Related product-matrix formulations have also been developed for polymers on finite graphs \cite{cometsMorenoRamirez2019}. For finite transverse geometries, Brunet and Derrida derived long-time free-energy cumulants on a cylinder \cite{brunetDerrida2000}, while more recent work obtained the corresponding cumulants and their large-size scaling for the KPZ equation on an interval \cite{barraquandLeDoussal2025}.

The connection between KPZ physics and random matrix theory (RMT) is now a cornerstone of the field. The appearance of Tracy-Widom laws in KPZ fluctuations was first understood through exact solutions of integrable growth models and exclusion processes, and through their mapping to determinantal/Pfaffian structures \cite{johansson2000shape,prahofer2000universal,baik2000limiting}. In this correspondence, the edge fluctuations of random matrices provide the universal limiting distributions (GUE, GOE, GSE), while KPZ models provide the dynamical physical realization \cite{tracywidom1994airy,tracywidom1996orthogonal,corwin2012kpz,takeuchi2018appetizer}. Subsequent exact solutions of the continuum KPZ equation and continuum directed polymer further cemented this link, showing explicitly how the one-point free-energy distribution crosses over to the Tracy-Widom laws in the long-time limit \cite{sasamoto2010exact,amir2011continuum}.

In the present work, we revisit the transfer-matrix formulation of a lattice DPRM and adopt a complementary viewpoint. We regard the time-ordered product of random transfer matrices
\begin{equation*}
W(t)=T(t)T(t-1)\cdots T(1)
\end{equation*}
as the central object. For a fixed realization of the bulk disorder, full-space point-to-point and point-to-line partition functions are different contractions of the same matrix $W(t)$. The stationary construction uses a Brownian-weighted initial vector, while the half-space construction modifies only the boundary entries of the same bulk transfer rule. Numerically, these choices yield the canonical one-point KPZ laws: Tracy--Widom GUE for point-to-point geometry, Tracy--Widom GOE for point-to-line geometry, Baik--Rains statistics for the Brownian-weighted line-to-point construction, and Tracy--Widom GSE for the absorbing half-space construction. In all four cases, we recover the expected $t^{1/3}$ fluctuation growth, and the centered, standardized distributions and low-order cumulants agree with the corresponding universal benchmarks.

This matrix-product perspective organizes the geometry-dependent KPZ subclasses in a single finite-dimensional transfer-matrix framework. The bulk stochastic evolution is kept fixed, while the observable is selected by endpoint contraction, initial boundary vector, or a local boundary modification.

At the same time, the transfer-matrix ensemble naturally contains observables that are not tied to canonical endpoint geometries. Beyond matrix elements, one may examine intrinsic spectral properties of $W(t)$, such as its leading eigenvalue. Motivated by the central role played by the edge eigenvalue in classical random matrix theory, we consider the growth of $\ln \lambda_1(t)$, where $\lambda_1(t)$ is the largest eigenvalue of $W(t)$. While $\ln \lambda_1(t)$ exhibits KPZ-like $t^{1/3}$ fluctuation growth over an intermediate time window, its standardized one-point distribution and low-order cumulants remain distinct from the canonical Tracy–Widom benchmarks within the numerically accessible regime. This suggests that the transfer-matrix ensemble may encode additional fluctuation structures beyond the geometry-selected KPZ subclasses.

The remainder of the paper is organized as follows. In Sec.~\ref{sec:dprm_tm} we introduce the lattice directed-polymer transfer-matrix formulation and define the random matrix-product ensemble studied here. In Sec.~\ref{sec:kpz_matrix_elements} we implement the four endpoint and boundary configurations and present numerical evidence for Tracy--Widom GUE, GOE, and GSE statistics, as well as Baik--Rains fluctuations in the stationary construction. In Sec.~\ref{sec:additional_observable} we turn to matrix-level observables beyond endpoint free energies, focusing on the leading eigenvalue of the transfer-matrix product. Finally, in Sec.~\ref{sec:discussion_outlook} we summarize the results and discuss future directions from the spectral perspective opened by this framework.

\section{Directed Polymer in Random Media and Transfer-Matrix Formulation}
\label{sec:dprm_tm}

\begin{figure}[t]
    \centering
    \includegraphics[width=1.0\linewidth]{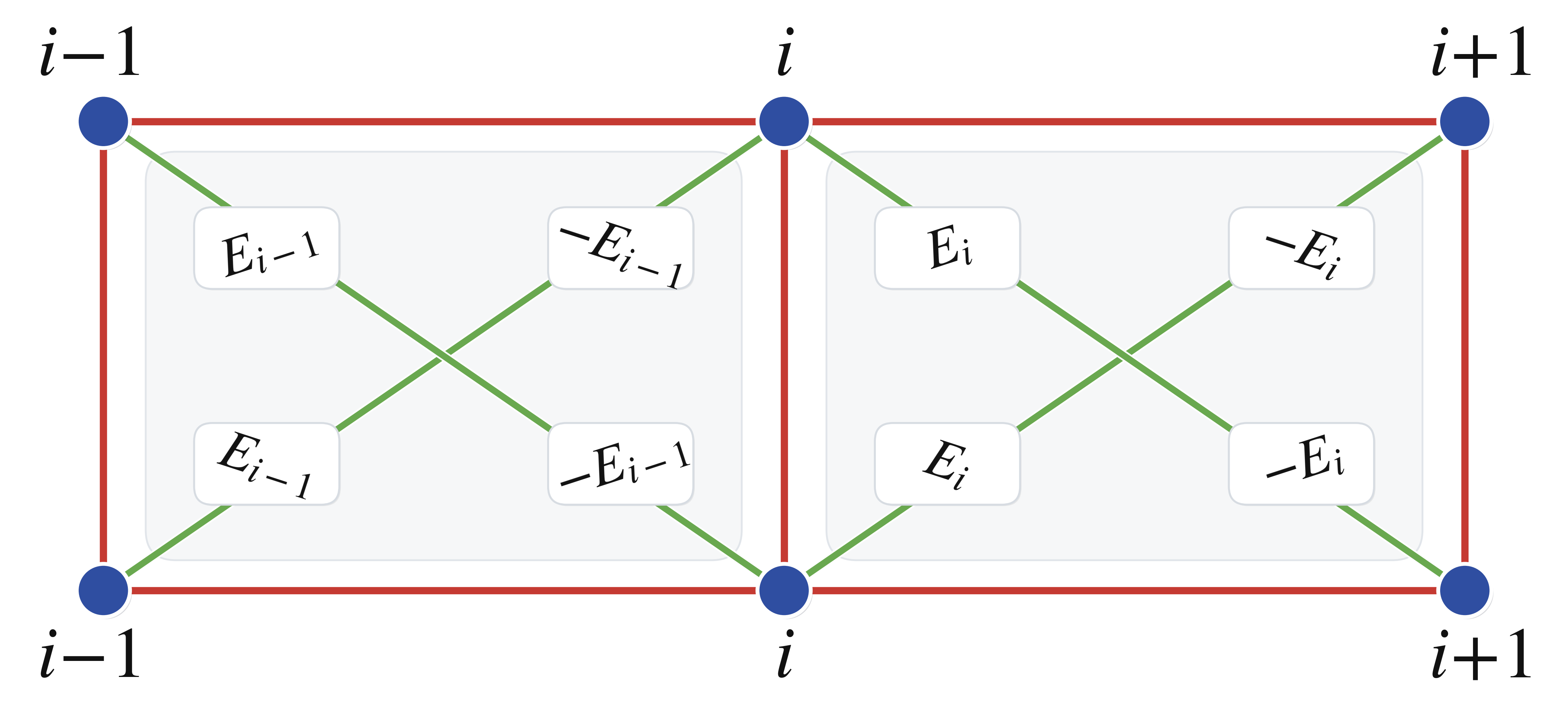}
    \caption{Illustration of the random energies on the green bonds in the 6-vertex model representation considered for the directed polymer in random media.}
    \label{fig:DPModel}
\end{figure}

\begin{figure*}[t]
\centering
\begin{tikzpicture}[x=0.55cm,y=0.48cm,>=stealth,font=\scriptsize]

\begin{scope}[xshift=0cm]
\node at (2,5.6) {(a) point-to-point};
\draw[->] (0,0) -- (4.5,0) node[right] {$t$};
\draw[->] (0,0) -- (0,4.8) node[above] {$x$};
\fill (0.3,2.2) circle (2pt);
\fill (4.1,2.2) circle (2pt);
\draw[thick] (0.3,2.2) .. controls (1.1,3.4) and (1.8,0.9) .. (2.5,2.6)
 .. controls (3.1,3.5) and (3.6,1.4) .. (4.1,2.2);
\node[align=center] at (2.2,-0.8) {$\langle x_f|W(t)|x_i\rangle$\\TW--GUE};
\end{scope}

\begin{scope}[xshift=4.2cm]
\node at (2,5.6) {(b) point-to-line};
\draw[->] (0,0) -- (4.5,0) node[right] {$t$};
\draw[->] (0,0) -- (0,4.8) node[above] {$x$};
\fill (0.3,2.2) circle (2pt);
\draw[thick] (4.05,0.6) -- (4.05,4.3);
\draw[thick] (0.3,2.2) .. controls (1.1,3.5) and (2.0,1.0) .. (3.0,2.8)
 .. controls (3.4,3.1) and (3.7,3.3) .. (4.05,3.4);
\node[align=center] at (2.2,-0.8) {$\sum_x\langle x|W(t)|x_i\rangle$\\TW--GOE};
\end{scope}

\begin{scope}[xshift=8.4cm]
\node at (2,5.6) {(c) Brownian line-to-point};
\draw[->] (0,0) -- (4.5,0) node[right] {$t$};
\draw[->] (0,0) -- (0,4.8) node[above] {$x$};
\draw[thick] (0.25,0.5) -- (0.25,4.3);
\draw[thick] (0.25,1.0) -- (0.55,1.3) -- (0.25,1.7) -- (0.55,2.0)
 -- (0.25,2.5) -- (0.55,3.0) -- (0.25,3.5);
\fill (4.05,2.2) circle (2pt);
\draw[thick] (0.4,3.0) .. controls (1.3,3.8) and (2.4,0.9) .. (4.05,2.2);
\node[align=center] at (2.2,-0.8) {$\langle x_f|W(t)|u\rangle$\\Baik--Rains};
\end{scope}

\begin{scope}[xshift=12.6cm]
\node at (2,5.6) {(d) half-space point-to-point};
\draw[->] (0,0) -- (4.5,0) node[right] {$t$};
\draw[->] (0,0) -- (0,4.8) node[above] {$x$};
\draw[very thick] (0.2,4.25) -- (4.1,4.25);
\node[above] at (2.2,4.25) {absorbing wall};
\fill (0.3,3.5) circle (2pt);
\fill (4.05,3.5) circle (2pt);
\draw[thick] (0.3,3.5) .. controls (1.0,2.0) and (1.8,3.8) .. (2.6,2.7)
 .. controls (3.2,2.1) and (3.6,3.1) .. (4.05,3.5);
\node[align=center] at (2.2,-0.8) {$\langle x_f|W_{\rm hs}(t)|x_i\rangle$\\TW--GSE};
\end{scope}
\end{tikzpicture}
\caption{Schematic of the four endpoint and boundary configurations used in
this work. The first two are different contractions of the same
full-space product $W(t)$. The stationary construction uses a
Brownian-weighted initial vector $|u\rangle$, while the half-space
construction uses the same bulk transfer rule with an absorbing
boundary.}
\label{fig:geometry_schematics}
\end{figure*}

We consider directed polymers in random media (DPRM), where the central object is the partition function obtained by summing over all directed paths connecting specified initial and final points. For a polymer starting at $(x_0,0)$ and ending at $(x,t)$, with $x_0,x\in[1,\ldots,N]$ and $t>0$, the partition function is defined as
\begin{equation}
Z(x,x_0,t) =
\sum_{\text{directed paths}}
\exp \left[ - \sum_{\alpha = 1}^{t} E_{x,x_0}^{\alpha} \right].
\label{eq:partition_fn_1_DPRM}
\end{equation}
Here, $\sum_{\alpha = 1}^{t} E_{x,x_0}^{\alpha}$ denotes the total random energy accumulated along each path, and the Boltzmann weight assigns larger statistical weight to lower-energy paths. The associated free energy is
\begin{equation}
F(x,x_0,t) = -\ln Z(x,x_0,t).
\end{equation}

A convenient way to evaluate the partition function is through a transfer-matrix formulation, which takes advantage of the directed nature of the paths. Denoting by $T(t)$ the transfer matrix at time slice $t$, the partition function satisfies the recursion relation
\begin{equation}
Z(x,x_0,t)
=\sum_{x'}\langle x \mid T(t) \mid x' \rangle \,Z(x',x_0,t-1)\,.
\end{equation}
It is therefore useful to introduce the ordered product of transfer matrices
\begin{equation}
W(t)=T(t)T(t-1)\cdots T(1).
\label{eq:product_matrix}
\end{equation}
so that the partition function can be written compactly as the matrix element
\begin{equation}\label{eq:FTM}
Z(x,x_0,t)
=\langle x \mid W(t) \mid x_0 \rangle\,.
\end{equation}
Equation~\eqref{eq:FTM} makes clear that full-space polymer observables with different endpoint conditions are matrix elements or contractions of the same product $W(t)$. In the stationary construction below, the same bulk product is contracted with a random Brownian initial vector, whereas in half space the transfer rule is modified only at the absorbing boundary. Thus the bulk random environment is common to all cases, while the geometry is encoded by the endpoint contraction, the boundary vector, or the boundary constraint.

In this work, we adapt a specific realization of DPRM motivated by a 6-vertex model in a random environment. As depicted in Fig.~\ref{fig:DPModel}, random energies are assigned to diagonal bonds of the model in a manner that leads to a particularly simple local transfer matrix. We adopt this choice because it provides an efficient and transparent transfer-matrix representation, while preserving the essential ingredients of the problem.
The corresponding tridiagonal transfer matrix at each time slice now takes the form
\begin{equation}
T(t^\prime) =
\begin{pmatrix}
M_{11}^{t^\prime} & 1 & 0 & 0 & \cdots \\
1 & M_{22}^{t^\prime} & 1 & 0 & \cdots \\
0 & 1 & M_{33}^{t^\prime} & 1 & \cdots \\
0 & 0 & 1 & M_{44}^{t^\prime} & \cdots \\
\vdots & \vdots & \vdots & \vdots & \ddots
\end{pmatrix},
\end{equation}
with time-dependent diagonal entries
\begin{equation}
M_{ii}^{t^\prime} =  e^{-2 E_{i-1}^{t^\prime}}+e^{2 E_{i}^{t^\prime}}.
\end{equation}
The random variables $E_i^{t'}$ are the (diagonal) bond energies of the corresponding vertex model (Fig.~\ref{fig:DPModel}) taken to be independent and identically distributed random variables drawn from a uniform distribution with mean $\mu$ and standard deviation $\sigma$,
\begin{equation}
E^{t^\prime}_i \in [\mu-\sqrt{3}\sigma,\mu+\sqrt{3}\sigma].
\end{equation}
At the boundaries, the diagonal entries of the transfer matrix are modified according to the vertex-model geometry. In particular, we impose
\begin{equation}
M_{11}^{t^\prime} = e^{-2 E_{0}^{t^\prime}}+e^{2 E_{1}^{t^\prime}},\qquad
M_{NN}^{t^\prime} = e^{-2 E_{N-1}^{t^\prime}}.
\end{equation}
These boundary terms complete the definition of the DPRM transfer-matrix ensemble considered in this work.

Transfer-matrix recursions have long been used to compute partition sums and finite-time free-energy distributions of lattice directed polymers~\cite{derridaGolinelli1990,halpinHealy1991}. The specific vertex-motivated kernel above is a particular microscopic realization studied here. To our knowledge, this precise kernel has not previously been used to organize the four one-point subclasses considered here.
It is local and of finite range: the bond variables $E_i^{t'}$ are independent between different time slices and are independently sampled in space, although the same $E_i^{t'}$ enters two neighboring diagonal elements and therefore induces a nearest-neighbor correlation between  matrix elements. The disorder is bounded and short ranged. These assumptions define the numerical model considered in this paper. Spatially long-range-correlated or heavy-tailed environments can lead to different scaling behavior~\cite{chuKardar2016correlated,gueudreEtAl2015heavy}. Temporally correlated environments also lie outside the independent-in-time ensemble considered here. While the transfer-matrix recursion can be applied to correlated environments,  the universality class studied below refers to the independent-in-time, short-range-disorder ensemble specified here.

Throughout this work, we focus on system sizes $N$ and evolution times $t$ chosen so as to probe the pre-saturation regime of KPZ scaling. For $1 \ll t \ll N^{3/2}$, finite-size effects are subdominant and free-energy fluctuations are expected to exhibit the characteristic $t^{1/3}$ growth associated with KPZ universality. The numerical results presented below are obtained within this scaling window.

\begin{figure}[t]
    \centering
    \includegraphics[width=1.0\linewidth]{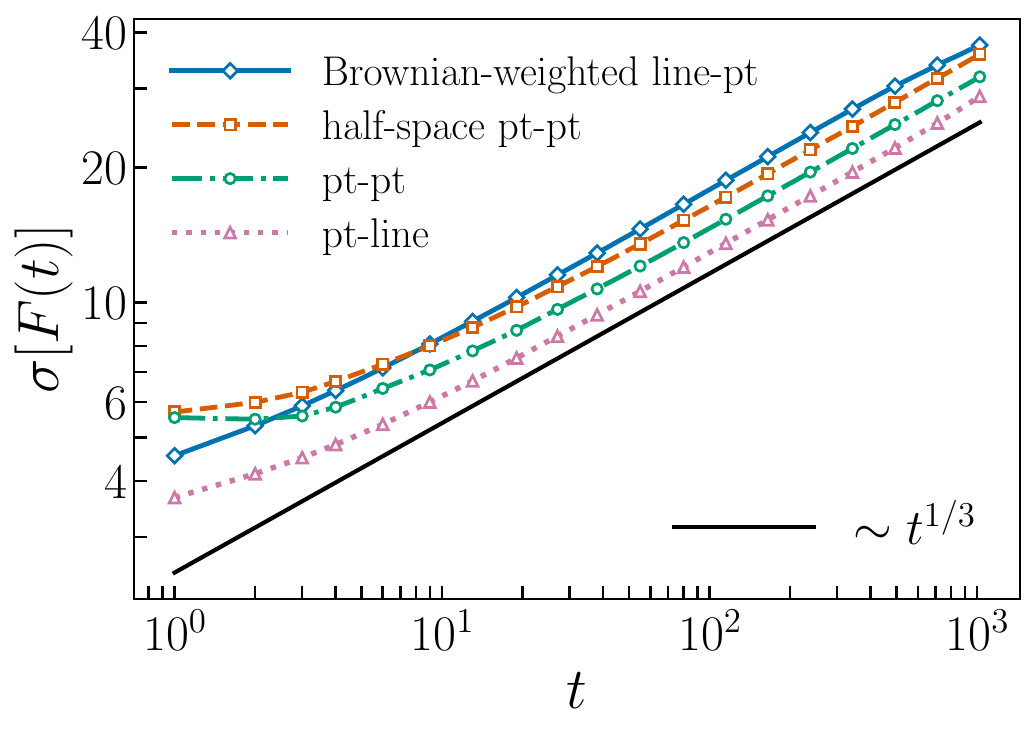}
   \caption{Standard deviation of the free energy, $\sigma[F(t)]$, as a function of time $t$ for the Brownian-weighted line-to-point, half-space point-to-point, point-to-point, and point-to-line boundary conditions, plotted on log-log scales. The black solid line indicates the power law $t^{1/3}$. The data are consistent with KPZ scaling of the free-energy fluctuations in all four cases. Model parameters are $N=128$, $\mu=0$, and $\sigma=3$, with evolution up to $t=1024$ over $10^6$ disorder realizations.}
    \label{fig:fluctuation_scaling}
\end{figure}

\section{KPZ One-Point Subclasses from Product Matrix Elements}
\label{sec:kpz_matrix_elements}

It is well established that, for DPRM in $1+1$ dimensions, endpoint geometry and boundary or initial data select distinct one-point KPZ fluctuation subclasses~\cite{prahofer2000universal,baik2000limiting,corwin2012kpz,takeuchi2018appetizer}. Figure~\ref{fig:geometry_schematics} summarizes the four configurations considered here. In the transfer-matrix language, point-to-point and point-to-line observables are two contractions of the same full-space product $W(t)$. The stationary construction supplements this product with a Brownian-weighted initial vector, while the half-space construction modifies only the transfer rule at the absorbing boundary.

Fixing both endpoints realizes the droplet geometry and gives the Tracy--Widom GUE law, whereas summing uniformly over one endpoint realizes the flat geometry and gives the Tracy--Widom GOE law~\cite{johansson2000shape,prahofer2000universal,calabreseLeDoussal2012flat}. A two-sided Brownian initial profile gives the stationary subclass and the Baik--Rains distribution~\cite{baik2000limiting,prahofer2000universal,barraquandLeDoussal2022steady,barraquandLeDoussal2023stationary}. In half space, an absorbing or sufficiently repulsive wall leads to the Tracy--Widom GSE subclass~\cite{sasamoto2004halfspace,imamura2004external,gueudreLeDoussal2012hardwall}.

In all of these subclasses, we characterize the fluctuations of the free energy through its standard deviation
\begin{equation}
\sigma[F(t)] = \sqrt{\langle F(t)^2 \rangle - \langle F(t) \rangle^2},
\label{eq:F_std}
\end{equation}
where $\langle \cdots \rangle$ denotes the disorder average. As shown in Fig.~\ref{fig:fluctuation_scaling}, we find that $\sigma[F(t)]$ exhibits a clear power-law growth in time in all four cases shown, consistent with the KPZ scaling form $\sigma[F(t)] \sim t^{1/3}$.

In the following, we present the full distributions of the standardized free energy,
\begin{equation}
\tilde{F}(t) = -\big(F(t)-\langle F(t)\rangle\big)/\sigma[F(t)],
\label{eq:normed_F}
\end{equation}
for each of these subclasses and compare them with the corresponding universal benchmark distributions.

The minus sign in Eq.~\eqref{eq:normed_F} follows from the convention $F=-\ln Z$. The KPZ height variable corresponding to the polymer partition function is $h=\ln Z=-F$; hence $\tilde F$ is the standardized height-like fluctuation. This convention gives the usual orientation of the Tracy--Widom and Baik--Rains laws used in the KPZ literature~\cite{sasamoto2010exact,amir2011continuum,corwin2012kpz}.

\subsection{Tracy-Widom GUE and GOE distributions}

\begin{figure}[t]
\centering
\includegraphics[width=1.0\linewidth]{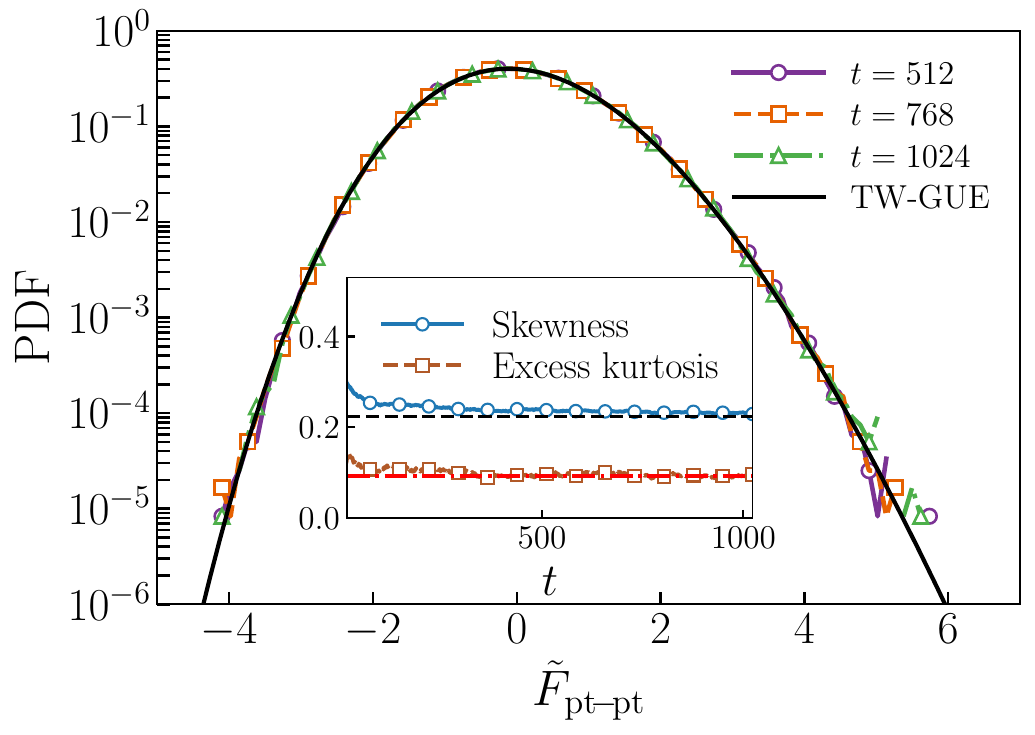}
\caption{Statistics of the standardized free energy for the point-to-point boundary condition. Main: Distribution of the standardized free energy $\tilde{F}_{\rm pt\text{-}pt}$ at different times ($t=512,768,1024$), together with the standardized TW-GUE distribution. Inset: Time evolution of the skewness and excess kurtosis of the standardized free energy. The black dashed and red dot-dashed lines denote the corresponding TW-GUE values, 0.22 and 0.09, respectively. Model parameters are $N=128$, $\mu=0$, $\sigma=3$, and $x_0=64$ with evolution up to $t=1024$ over $10^6$ disorder realizations.}
\label{fig:stats_pt_pt_matrix_elements}
\end{figure}

We first consider two basic contractions of the product matrix $W(t)$ corresponding to fixed and partially summed endpoints:
\begin{equation}
\begin{cases}
    F_{\rm pt\text{-}pt}(t) = -\ln \langle x_0 |W(t)| x_0 \rangle,
    & \text{for point-to-point,} \\
    F_{\rm pt\text{-}line}(t) = -\ln\sum_x \langle x | W(t)| x_0 \rangle,
    & \text{for point-to-line,}
\end{cases}
\label{eq:Wt-free-energy-def}
\end{equation}
with fixed $x_0$. In the first case, both endpoints are fixed, while in the second case, the final endpoint is summed uniformly over all transverse positions. These two contractions are expected to probe the droplet (TW-GUE) and flat (TW-GOE) subclasses, respectively~\cite{johansson2000shape,prahofer2000universal,calabreseLeDoussal2012flat}.

We present the free-energy statistics for the point-to-point boundary condition in Fig.~\ref{fig:stats_pt_pt_matrix_elements}. The standardized distribution at late time is well described by the Tracy--Widom GUE benchmark. This identification is further supported by the time evolution of the skewness and excess kurtosis, which approach the corresponding TW-GUE values within the accessible time range.

\begin{figure}
\centering
\includegraphics[width=1.0\linewidth]{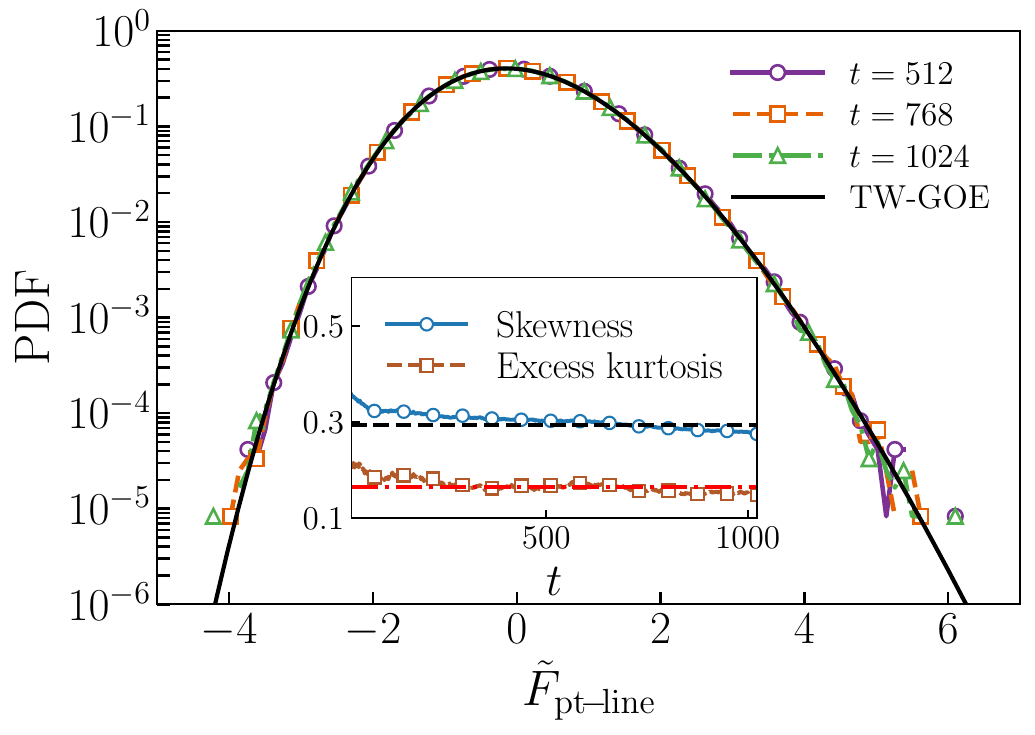}
\caption{Statistics of the standardized free energy for the point-to-line boundary condition. Main: Distribution of the standardized free energy $\tilde{F}_{\rm pt\text{-}line}$ at different times ($t=512,768,1024$), together with the standardized TW-GOE distribution. Inset: Time evolution of the skewness and excess kurtosis of the standardized free energy. The black dashed and red dot-dashed lines denote the corresponding TW-GOE values, 0.29 and 0.17, respectively. Model parameters are $N=128$, $\mu=0$, $\sigma=3$ and $x_0=64$ with evolution up to $t=1024$ over $10^6$ disorder realizations.}
\label{fig:stats_pt_line_matrix_elements}
\end{figure}

We next present the free-energy statistics for the point-to-line boundary condition in Fig.~\ref{fig:stats_pt_line_matrix_elements}. In this case the standardized distribution is well described by the Tracy--Widom GOE benchmark, as expected for the flat subclass. This identification is again supported by the corresponding skewness and excess kurtosis, which evolve toward the TW-GOE values at late times.

\subsection{Tracy-Widom GSE distribution}

\begin{figure}
\centering
\includegraphics[width=1.0\linewidth]{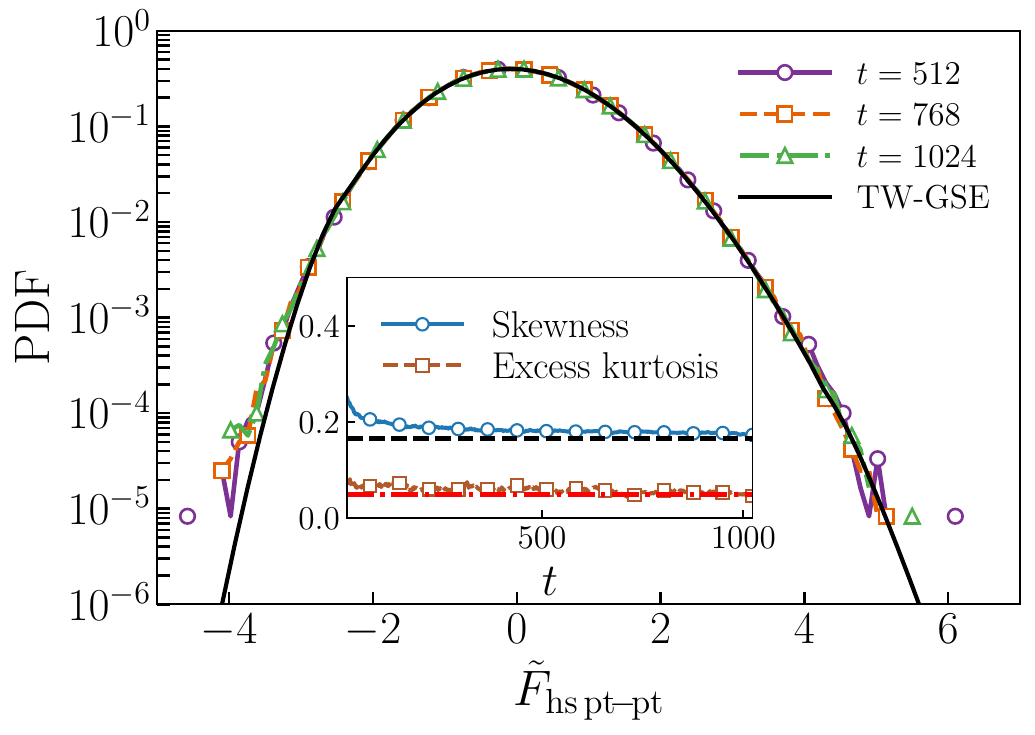}
\caption{Statistics of the standardized free energy for the point-to-point boundary condition in half space. Main: Distribution of the standardized free energy $\tilde{F}_{\rm hs\, pt\text{-}pt}$ at different times ($t=512,768,1024$), together with the standardized TW-GSE distribution. Inset: Time evolution of the skewness and excess kurtosis of the standardized free energy. The black dashed and red dot-dashed lines denote the corresponding TW-GSE values, 0.16 and 0.04, respectively. Model parameters are $N=128$, $\mu=0$, $\sigma=3$ and $x_0=127$ with evolution up to $t=1024$ over $10^6$ disorder realizations.}
\label{fig:stats_half_space_pt_pt_matrix_elements}
\end{figure}

We next introduce an absorbing boundary while leaving the bulk transfer rule and bulk disorder distribution unchanged. The polymer is restricted to the half-space adjacent to the last site $x=N$ and starts and ends at $x_0=N-1$, as sketched in Fig.~\ref{fig:geometry_schematics}(d). A trajectory that enters the wall state is absorbed and cannot return to the active half-space. With the convention that $\langle x|T(t')|x'\rangle$ propagates a source position $x'$ to a destination position $x$, this condition is imposed by setting to zero all matrix elements that propagate the wall state back into the active half-space:
\begin{equation}
\forall\,x'\in\{1,\ldots,N-1\},\,t':
\qquad
\langle x'|T_{\rm hs}(t')|N\rangle=0.
\label{eq:absorbing_wall}
\end{equation}
Thus no propagation can return from the absorbing state to the active half-space. This is the standard hard-wall geometry associated with the GSE half-space subclass~\cite{sasamoto2004halfspace,imamura2004external,gueudreLeDoussal2012hardwall}.
We denote the corresponding half-space product by
\begin{equation}
W_{\rm hs}(t)=T_{\rm hs}(t)T_{\rm hs}(t-1)\cdots T_{\rm hs}(1).
\end{equation}
With this boundary condition, the half-space point-to-point free energy is
\begin{equation}
F_{\rm hs\,pt\text{-}pt}(t)
=-\ln\langle x_0|W_{\rm hs}(t)|x_0\rangle,
\end{equation}
with $x_0=N-1$ and the absorbing-wall condition in Eq.~\eqref{eq:absorbing_wall}.

For numerical convenience, we retain $x=N$ as an auxiliary absorbing state. The polymer may enter this state from $x=N-1$, but the zero upper-right matrix element prevents a transition back to $x=N-1$. Accordingly, the block of each one-step transfer matrix acting on $\{|N-1\rangle,|N\rangle\}$ is triangular:
\begin{equation}
T_{{\rm hs},\{N-1,N\}}(t')=
\begin{pmatrix}
M^{t'}_{N-1} & 0\\
1 & M^{t'}_{N}
\end{pmatrix}.
\label{eq:jordan_block}
\end{equation}
The lower-left element permits $N-1\to N$, while the vanishing upper-right element forbids $N\to N-1$. The diagonal element $M_N^{t'}$ only propagates weight that has already entered the absorbing state and cannot contribute to the return matrix element $\langle x_0|W_{\rm hs}(t)|x_0\rangle$. Consequently, every trajectory touching the wall is excluded from this point-to-point partition sum.

For the bond-disorder transfer matrix and hard-wall boundary
considered here, an equivalent half-space implementation is obtained
by placing the auxiliary absorbing state at $x=N+1$. In this
representation, the active $N\times N$ block is precisely the
original matrix $T(t)$, and the half-space partition function is
represented by the edge-to-edge element
$\langle N|W(t)|N\rangle$. Any path entering the auxiliary state
cannot return and therefore cannot contribute to this matrix element.
This identification with the original matrix relies on the
microscopic boundary rule of the present model; more general
half-space disorder or boundary interactions may also modify the
active edge weights. The absorbing wall considered here corresponds
to the unbound hard-wall regime, for which GSE fluctuations are
expected.

The free-energy statistics obtained with the absorbing-state
implementation are shown in
Fig.~\ref{fig:stats_half_space_pt_pt_matrix_elements}. We find that
the standardized free-energy distribution is well described by the
Tracy--Widom GSE benchmark. This identification is further supported
by the time evolution of the skewness and excess kurtosis, which move
toward the corresponding TW-GSE values at late times.

\subsection{Baik-Rains distribution}

\begin{figure}
\centering
\includegraphics[width=1.0\linewidth]{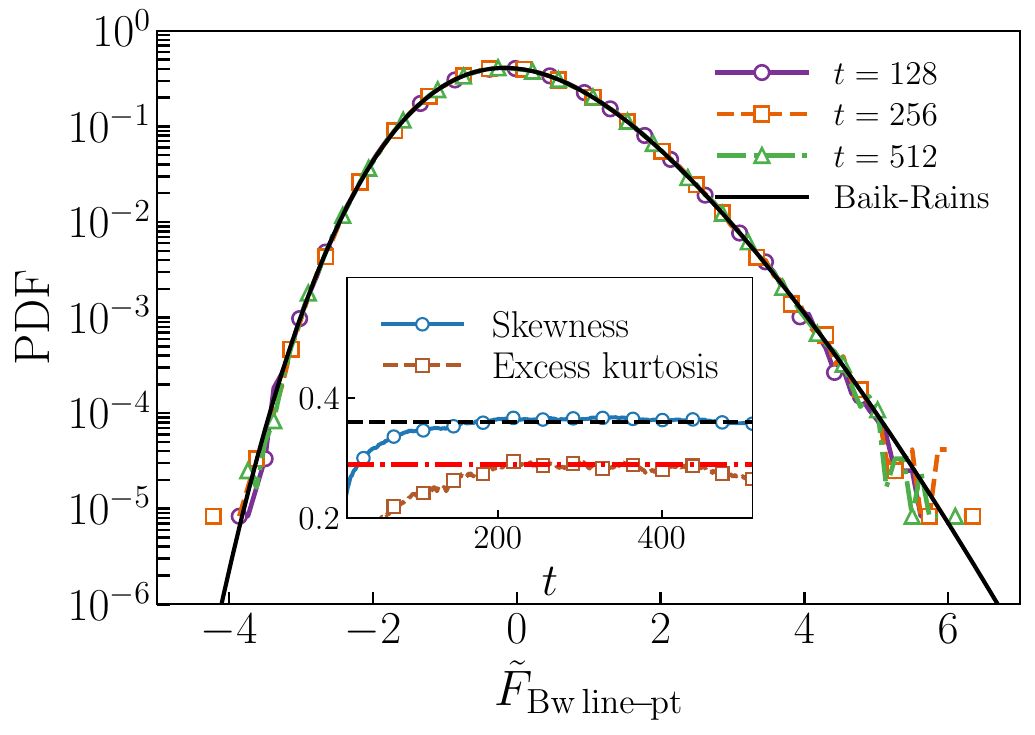}
\caption{Statistics of the standardized free energy for the Brownian-weighted line-to-point boundary condition. Main: Distribution of the standardized free energy $\tilde{F}_{\rm Bw\, line\text{-}pt}$ at different times ($t=128,256,512$), together with the standardized Baik-Rains distribution. Inset: Time evolution of the skewness and excess kurtosis of the standardized free energy. The black dashed and red dot-dashed lines denote the corresponding Baik-Rains values, 0.36 and 0.29, respectively. Model parameters are $N=128$, $\mu=0$, $\sigma=3$, $x_0=64$ and $\sigma_B=6.2$ with evolution up to $t=512$ over $10^6$ disorder realizations.}
\label{fig:stats_BW_line_pt_matrix_elements}
\end{figure}

We now turn to a construction aimed at probing the stationary KPZ subclass within the same transfer-matrix framework. On the full line, the stationary KPZ subclass is obtained from a two-sided Brownian initial profile; its height increments remain stationary and its one-point fluctuations converge to the Baik--Rains distribution~\cite{baik2000limiting,prahofer2000universal}. On finite intervals, exact stationary measures can depend nontrivially on the boundary conditions, as shown in recent studies of the KPZ equation and directed polymers on an interval~\cite{barraquandLeDoussal2022steady,barraquandLeDoussal2023stationary}.

To approximate this setting in the present discrete transfer-matrix formulation, we introduce a Brownian-weighted initial vector $|u\rangle$ with components
\begin{equation}
u(x)=e^{B(x)},
\end{equation}
where $B(x)$ is a discrete two-sided random walk referenced to a central site and fixed by the convention $B(x_{\rm mid})=0$. This removes an arbitrary additive constant and defines a random initial free-energy profile. Brownian-type stationary structures also arise in extensions to several nonintersecting directed polymers~\cite{barraquandLeDoussal2023nonintersecting}.
The corresponding free energy is obtained from the contraction
\begin{eqnarray}
F_{\rm Bw\, line\text{-}pt}(t)&=&-\ln\sum_x e^{B(x)}\langle x_0|W(t)|x\rangle \nonumber\\
&=&-\ln\langle x_0|W(t)|u\rangle,
\end{eqnarray}
which represents a Brownian-weighted line-to-point observable from the same product matrix $W(t)$. In other words, the stationary-type boundary condition is implemented by summing over all starting positions with random initial weights $e^{B(x)}$, while keeping the bulk disorder ensemble unchanged. After centering and rescaling, the resulting one-point free-energy distribution is then compared with the Baik--Rains benchmark.

In the discrete transfer-matrix realization considered here, the exact stationary measure of the microscopic dynamics is not known analytically and may depend on the disorder distribution as well as on the update rule. We therefore treat the standard deviation parameter  $\sigma_B$  of the Brownian increments as an effective tuning parameter. Specifically, we determine $\sigma_B$  empirically by requiring that the late-time skewness and excess kurtosis of the standardized free-energy fluctuations approach the Baik–Rains benchmark values as closely as possible within the accessible time window.

It is important to emphasize that this Brownian-weighted initialization should be viewed as an effective approximation to the stationary subclass rather than an exact invariant measure of the discrete dynamics. Finite-size and finite-time corrections can affect one-point and two-point observables differently, and the optimal value of $\sigma_B$ for matching one-point cumulants need not coincide with that inferred from spatial roughness diagnostics. Within the present system sizes and evolution times, we therefore focus on one-point fluctuation statistics as the primary diagnostic of the stationary subclass.

The numerical results are shown in Fig.~\ref{fig:stats_BW_line_pt_matrix_elements}.

After centering and rescaling, the one-point free-energy distribution exhibits good agreement with the Baik–Rains benchmark over the accessible time range, and the low-order cumulants approach the expected universal values. This indicates that, within the unified transfer-matrix ensemble, the stationary KPZ subclass can be effectively realized through a Brownian-weighted contraction of the same product matrix $W(t)$.

\section{Matrix--level Observables beyond endpoint geometries}
\label{sec:additional_observable}

The contractions considered in Sec. III probe DPRM free energies associated with specific endpoint geometries. However, the transfer-matrix product $W(t)$ contains more information than its matrix elements alone. It is therefore natural to ask whether intrinsic matrix-level observables, independent of canonical boundary geometries, also exhibit KPZ-like fluctuation behavior.
In this section, we focus on one such observable: the largest eigenvalue of the product matrix $W(t)$.

Products of random matrices and their Lyapunov spectra have a long history in probability, dynamical systems, and disordered statistical mechanics~\cite{furstenbergKesten1960,oseledets1968,hennion1997}. Of particular relevance here, Eckmann and Wayne established an explicit connection between the largest Lyapunov exponent of a positive random-matrix product and a directed polymer in a random environment~\cite{eckmannWayne1989}. Related finite-volume directed-polymer studies have determined long-time free-energy cumulants on a cylinder and on an interval~\cite{brunetDerrida2000,barraquandLeDoussal2025}. Product-matrix formulations and long-time asymptotics have also been studied for directed polymers on the complete graph~\cite{cometsMorenoRamirez2019}. These works motivate examining spectral growth observables of the structured product $W(t)$ in addition to its geometry-dependent matrix elements.

\subsection{Largest Eigenvalue}\label{subsec:largest_eigenvalue}

Let $\{\lambda_i(t)\}_{i=1}^N$ denote the eigenvalues of the product matrix $W(t)$, ordered by decreasing modulus,
\begin{equation}
|\lambda_1(t)| \geq |\lambda_2(t)| \geq \cdots \geq |\lambda_N(t)|.
\end{equation}
Each transfer matrix is nonnegative and irreducible, with positive diagonal and nearest-neighbor entries. The product $W(t)$ is likewise nonnegative and irreducible.
By the Perron--Frobenius theorem, its spectral radius is therefore a real, positive, and simple eigenvalue. We denote this leading eigenvalue by $\lambda_1(t)$. This makes
\begin{equation}
F_1(t)=\ln \lambda_1(t)
\end{equation}
a natural spectral growth observable to study.

\begin{figure}
    \centering
    \includegraphics[width=1.0\linewidth]{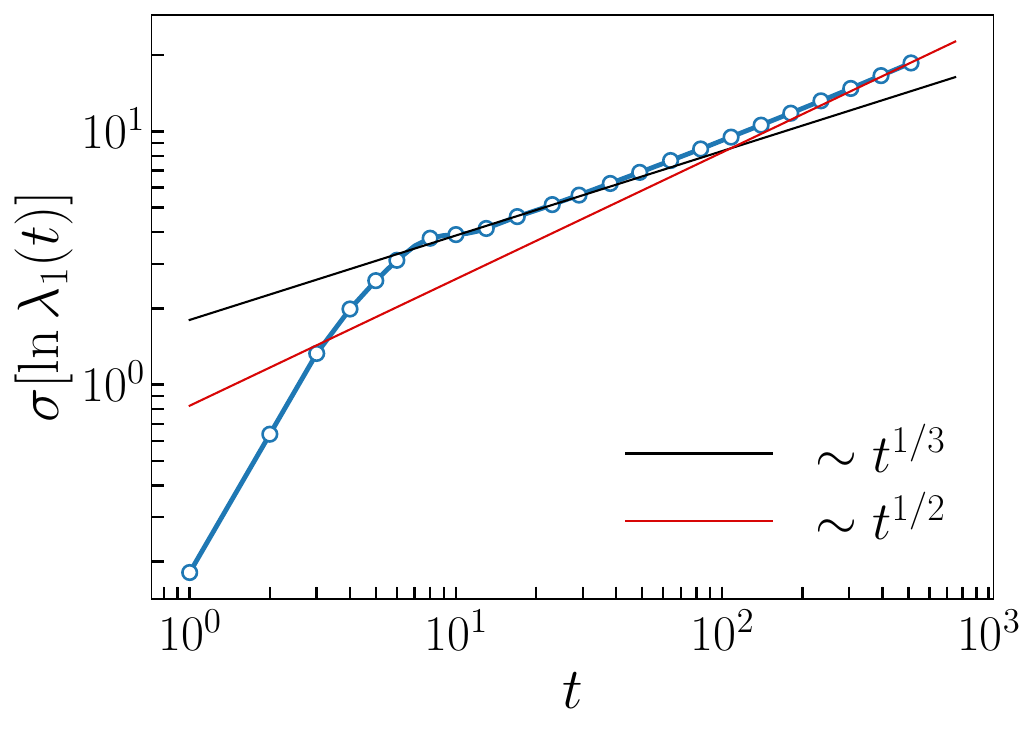}
    \caption{Standard deviation of the logarithmic leading eigenvalue, $\sigma[\ln \lambda_1(t)]$, as a function of time $t$ on log-log scales. The solid guide lines indicate the power laws $t^{1/3}$ (black) and $t^{1/2}$ (red). The data display an intermediate regime compatible with KPZ-like $t^{1/3}$ growth, followed at later times by a crossover toward a steeper scaling. Model parameters are $N=128$, $\mu=0$ and $\sigma=3$ with evolution up to $t=512$ over $10^6$ disorder realizations.}
    \label{fig:std-of-largest-eigenvalue}
\end{figure}

Unlike the canonical point-to-point or point-to-line free energies, $\ln\lambda_1(t)$ is not associated with a specific endpoint geometry. It therefore provides an example of a matrix-level observable intrinsic to the ensemble defined by $W(t)$.
To assess whether $\ln\lambda_1(t)$ displays KPZ-like behavior, we examine the growth of its fluctuations. Over an intermediate time window, we observe that the standard deviation grows approximately as
\begin{equation}
\sigma[\ln\lambda_1(t)] \sim t^{1/3}.
\end{equation}
This behavior is consistent with KPZ scaling, as shown in Fig.~\ref{fig:std-of-largest-eigenvalue}. This scaling regime is restricted in time and limited by finite-$N$ effects; beyond a crossover scale, deviations become visible.

We are not aware of an analytical prediction for the standardized distribution of $\ln\lambda_1(t)$ in the intermediate regime observed here. Existing results concern related but different limits. For fixed transverse size and very long times, standard results for positive random-matrix products imply that the logarithm of a fixed matrix element and the logarithm of the leading eigenvalue of the same finite-strip product are governed by the same leading growth rate~\cite{eckmannWayne1989,hennion1997}. For the boundary rule of the present model, the half-space
observable may equivalently be represented as an edge matrix
element of this same finite-strip product. This long-time relation does not determine the standardized distributions and, in particular, does not imply a Baik--Rains law for either observable. Long-time free-energy cumulants have been calculated for directed polymers on a cylinder~\cite{brunetDerrida2000} and for the KPZ equation on an interval~\cite{barraquandLeDoussal2025}. These fixed-size, long-time results do not predict the distribution of $\ln\lambda_1(t)$ in the intermediate pre-saturation regime studied here.

\begin{figure}
    \centering \includegraphics[width=1.0\linewidth]{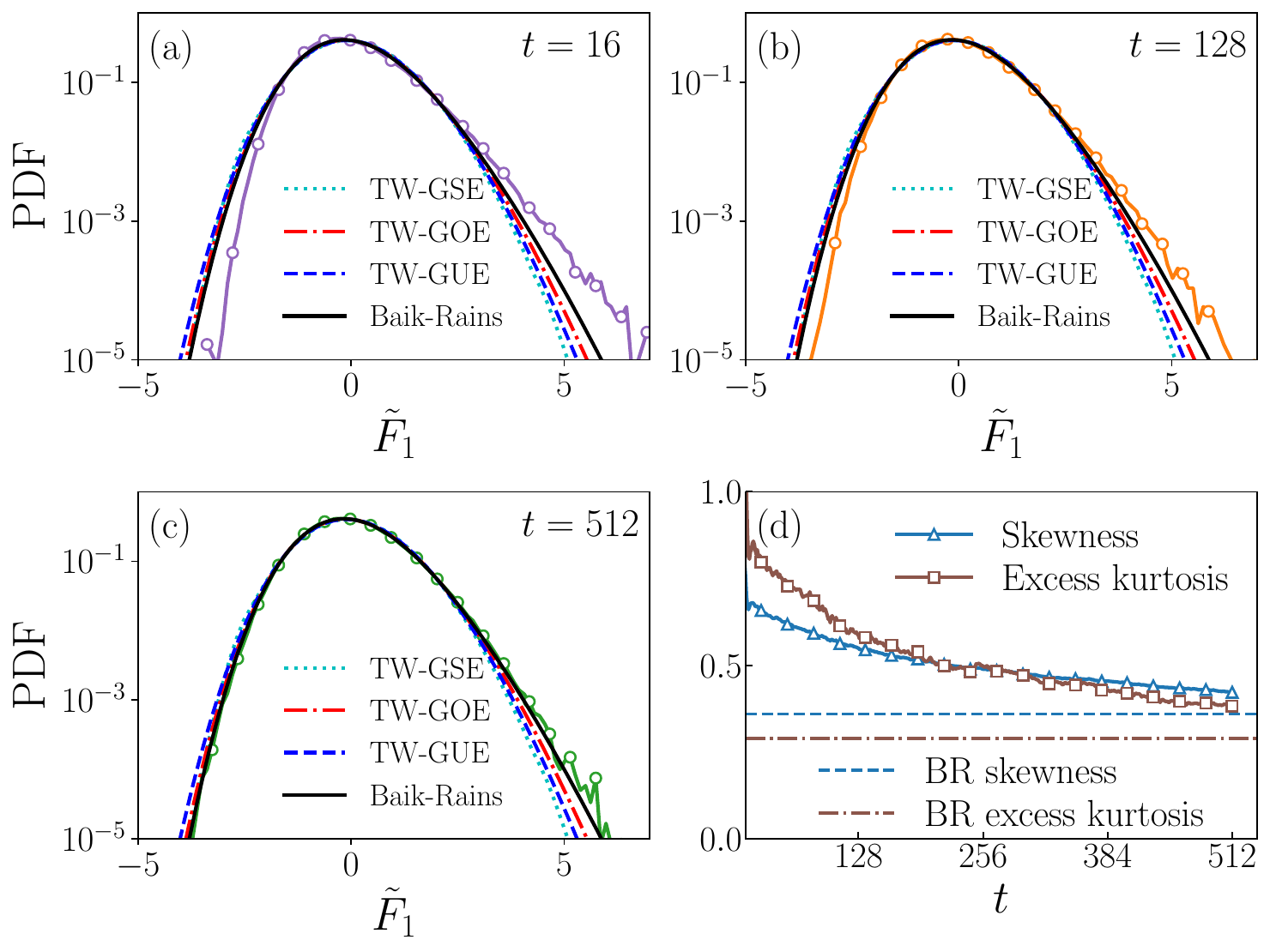}
\caption{Statistics of the standardized logarithmic leading eigenvalue $\tilde{F}_1(t)$. Panels (a)--(c) show the distribution of $\tilde{F}_1(t)$ at times $t=16,128,512$, compared with the standardized TW-GSE, TW-GOE, TW-GUE, and Baik--Rains benchmark distributions. Panel (d) shows the time evolution of the skewness and excess kurtosis of $\tilde{F}_1(t)$. The horizontal dashed and dot-dashed lines indicate the Baik--Rains skewness and excess kurtosis, respectively, shown only as reference values for comparison. Among these canonical laws, Baik--Rains provides the closest overall reference to the numerical data over the accessible range and is therefore included as a guide to the eye, rather than as an identified limiting law. Model parameters are $N=128$, $\mu=0$, and $\sigma=3$, with evolution up to $t=512$ over $10^6$ disorder realizations.}
\label{fig:distribution_largest_eigenvalues}
\end{figure}

We also examine the full distribution of the standardized $\ln\lambda_1(t)$ at different times and compute its skewness and excess kurtosis, presented in Fig.~\ref{fig:distribution_largest_eigenvalues}. We define the standardized logarithmic leading eigenvalue as
\begin{equation}
\tilde F_1(t)=
\frac{\ln\lambda_1(t)-\langle\ln\lambda_1(t)\rangle}
{\sigma[\ln\lambda_1(t)]}.
\end{equation} Within the accessible time range, these cumulants remain systematically distinct from the canonical Tracy–Widom GUE and GOE benchmark values. In particular, no clear convergence toward known TW cumulants is observed over the simulated regime.

Taken together, these results indicate that $\ln\lambda_1(t)$ exhibits KPZ-like $t^{1/3}$ fluctuation growth over an intermediate window, while its one-point statistics do not coincide with geometry--selected Tracy--Widom or Baik--Rains subclasses discussed in Sec.~\ref{sec:kpz_matrix_elements}.

\subsection{Spectral Perspective}\label{spectral_perspective}

The leading eigenvalue provides only the simplest example of an intrinsic matrix-level observable. More generally, the full spectrum of $W(t)$ encodes information about the structure of the random matrix product associated with DPRM evolution.
In contrast to the endpoint free energies, which are selected by specific boundary contractions, spectral observables are determined by the internal structure of $W(t)$ itself. The transfer-matrix viewpoint, therefore, suggests a broader class of fluctuation observables derived from the same disorder ensemble. While the present work focuses on the largest eigenvalue as a representative, a systematic study of the spectral statistics of $W(t)$ is desirable, but lies beyond the scope of this paper.

\section{Discussion and Outlook}
\label{sec:discussion_outlook}

We have shown numerically that a common lattice DPRM transfer-matrix framework provides a compact realization of the canonical $(1+1)$-dimensional KPZ one-point subclass structure. The central full-space object is the matrix
$W(t)=T(t)T(t-1)\cdots T(1)$.
Point-to-point and point-to-line partition sums are obtained by different endpoint contractions of this product. The stationary construction uses a Brownian-weighted initial vector, while the half-space construction changes only the transfer rule at the absorbing boundary. Thus all four cases share the same bulk disorder ensemble and local bulk dynamics, with geometry encoded by endpoint contractions, boundary data, or the boundary constraint.

For the canonical endpoint observables, our numerics reproduce the expected $t^{1/3}$ free-energy fluctuation scaling and yield one-point distributions that are in good agreement with the corresponding universal benchmarks. The droplet and flat contractions approach Tracy--Widom GUE and GOE, respectively; the half-space construction yields behavior consistent with Tracy–Widom GSE; and a Brownian-weighted contraction provides an effective realization of the Baik–Rains stationary subclass.
Taken together, these results show that geometry-dependent KPZ one-point subclasses can be organized directly at the level of a finite-dimensional matrix--product ensemble.

Beyond these canonical projections, the transfer-matrix product $W(t)$ naturally gives access to intrinsic matrix-level observables. As a representative example, we examined the logarithm of the leading eigenvalue $\ln\lambda_1(t)$; within the numerically accessible regime, this spectral observable exhibits KPZ-like  $t^{1/3}$  fluctuation growth over an intermediate time window. At the same time, its standardized distribution and cumulants remain distinct from the known Tracy–Widom and Baik–Rains subclasses over the simulated range. Thus, while the present results do not establish an asymptotic limiting law for this observable, they indicate that the unified transfer-matrix ensemble contains fluctuation structures not directly tied to the standard geometry-selected subclasses.

This perspective suggests a broader interpretation of the KPZ one-point structure. The familiar Tracy--Widom and Baik--Rains laws are selected within a common bulk transfer-matrix framework by endpoint contractions and boundary data. The transfer-matrix formulation therefore places the canonical KPZ subclasses alongside additional matrix-derived observables whose statistical properties remain to be explored.

Several directions follow naturally from the present viewpoint. On the numerical side, larger system sizes and longer times would clarify the intrinsic spectral observables such as $\ln\lambda_1(t)$. On the conceptual side, it would be interesting to understand more systematically how spectral observables of $W(t)$ relate to the other DPRM configurations, and whether additional matrix-level quantities admit universal descriptions. In particular, the fixed-width long-time cumulant results for polymers on cylinders and intervals~\cite{brunetDerrida2000,barraquandLeDoussal2025} provide a useful reference point, but a theory for the intermediate pre-saturation eigenvalue distribution reported here remains open. More broadly, the transfer-matrix approach opens a route to studying DPRM and KPZ fluctuations through structured random matrix products, complementing integrable and continuum formulations.

In summary, a common ensemble of finite-dimensional DPRM transfer matrices, supplemented by the appropriate endpoint contractions and boundary data, provides a unified framework for the canonical geometry-dependent KPZ one-point fluctuation laws and suggests a wider class of matrix-level observables beyond the canonical Tracy--Widom subclasses.

\section*{Acknowledgements} S.M. thanks Jonas Karcher for helpful discussions. This work was funded by the Deutsche Forschungsgemeinschaft (DFG, German Research Foundation) under Project No.~557852701 (A.A.S.). The study was also supported by the Advanced Study Group ``Strongly Correlated Extreme Fluctuations'' at the Max Planck Institute for the Physics of Complex Systems, Dresden (2024/25) \cite{pks_asg2024}, and the NSF through Grant No. DMR-2218849 (M.K.).

\section*{DATA AVAILABILITY}
The data that support the findings of this article are not
publicly available upon publication because it is not technically feasible and/or the cost of preparing, depositing, and
hosting the data would be prohibitive within the terms of this
research project. The data are available from the authors upon
reasonable request.

\bibliography{references}

\end{document}